\documentclass[a4paper]{article}
\usepackage{amsmath,amssymb,amsthm}
\usepackage{colortbl}
\usepackage{graphics}
\usepackage[margin=2.5cm]{geometry}
\usepackage[usenames,dvipsnames]{xcolor}
\usepackage[colorlinks=true,linkcolor=blue,citecolor=red,urlcolor=Violet,bookmarksdepth=subsection,final]{hyperref}

  \newcommand{\eps}{\varepsilon}
  \newcommand{\bnabla}{\bar{\nabla}}
\renewcommand{\d}{\mathrm{d}}
	\newcommand{\g}{\mathfrak{g}}
  \newcommand{\Lie}{\mathcal{L}}
\renewcommand{\u}[1]{\underline{#1}}

\title{A polynomial action for gravity with matter, gauge-fixing and ghosts}
\author{Igor Khavkine\\
	Department of Mathematics and TIFPA-INFN,\\
	University of Trento,\\
	Via Sommarive, 14, 38123 Povo, ITALY\\
	E-mail address: igor.khavkine@unitn.it
}
\date{\today}

\begin{document}
\maketitle

\begin{abstract}
We give a variational formulation of General Relativity, with coupling
to a cosmological constant, to scalar fields, to vector fields and to
spinor fields (all with possible mass and interaction terms). Among the
matter fields, we include ghosts corresponding to diffeomorphism and
Yang-Mills gauge symmetries, with kinetic terms given by gauge fixing
conditions leading to hyperbolic equations of motion for all fields. The
distinguishing characteristic of our Lagrangian density is its
polynomiality, in all dynamical fields and Lagrange multipliers, and its
validity for any number of spacetime dimensions.
\end{abstract}

\textbf{Keywords:}
general relativity, variational principle, polynomial,
Einstein-Palatini, scalar field, vector field, spinor field.

\textbf{MSC2010:}
58E30, 49S05, 83C05, 26C05

\vspace{2ex}

Since the introduction of the Einstein-Hilbert variational principle for
the equations of General Relativity, many alternative Lagrangians have
been proposed, based on different choices of variables and with
different algebraic or geometric structure. As long as an alternative
Lagrangian can be reduced to the Einstein-Hilbert one by an algebraic
redefinition or elimination of variables (that is, not involving the
resolution of any differential equations), the alternative choice is
largely equivalent to the original one. However, these alternative
formulations may be useful for particular purposes, like development of
geometric intuition, simplification of some calculations, non-linear
deformationor coupling to other fields.

The Einstein-Hilbert Lagrangian (an $n$-form on an $n$-dimensional
manifold)~\cite{MTW},
\begin{equation}\label{eq:EH}
	L_{\u a}[g] = v_{\u a}[g] g^{bc} R_{bc}[g] ,
\end{equation}
is a function of the metric tensor $g_{bc}$, with $g^{bc}$ the inverse
metric, $v_{\u a}[g]$ the metric volume form and $R_{bc}[g]$ the Ricci
tensor. We are using a compressed index notation for $n$-forms, where
$v_{\u a} = v_{a_1 \cdots a_n}$. When compared to other Lagrangians used
in fundamental physics, like the scalar (Klein-Gordon), vector (Maxwell
or Yang-Mills), spinor (Dirac), etc.\ Lagrangians, a striking
distinction of the Einstein-Hilbert Lagrangian is that it is
non-polynomial in the metric $g_{bc}$, its dynamical field.

While not essential from the physical point of view, polynomiality is a
convenient technical property. The main simplification it brings is at
the level of perturbation theory, either classical or quantum. When
writing the metric $g = \bar{g} + \eps h_1 + \eps^2 h_2 + \cdots$ as an
additive perturbation of a background metric $\bar{g}$, the equations
for $g$ give rise to a hierarchical family of linear equations for each
of the $h_i$, with inhomogeneous sources depending polynomially on the
$h_{j<i}$. For equations obtained from a non-polynomial Lagrangian
like~\eqref{eq:EH}, at each order of perturbation theory, the order of
the polynomial dependence of the inhomogeneous sources on the $h_{j<i}$
terms will increase. This polynomial order stays bounded only if the
original Lagrangian was itself polynomial. In perturbative quantum field
theory, we can analogously say that a non-polynomial Lagrangian gives
rise to new and higher order \emph{Feynman diagram
vertices}~\cite{Weinberg} at each perturbative order. Thus, the
polynomiality of the Lagrangian may significantly simplify the algebraic
structure of perturbation theory, at least at higher orders.

A polynomial Lagrangian obviously gives rise to polynomial
Euler-Lagrange equations. Another advantage of polynomiality shows
itself at the level of the formal analysis of Euler-Lagrange equations
as partial differential equations. Any partial differential equation can
be geometrically and invariantly represented as an equation on the space
of jets of the dynamical fields (the dependent variables) or,
equivalently, the subspace of the jet space given by the vanishing locus
of these equations~\cite{Seiler}. When the equations are polynomial, this
locus is an algebraic subvariety of the jet space, the \emph{PDE
subvariety}. In the general non-polynomial case, it is at best a smooth
manifold or only a stratified manifold, when singularities are present.
Understanding the geometry of the PDE subvariety is crucial for
identifying symmetries and conservation laws, identifying integrability
conditions, prolonging the equations to involution, analyzing the
singularities of solutions, etc. Thus, the powerful machinery of
algebraic geometry, dedicated to the analysis of the geometric
properties of algebraic varieties, may be brought to bear on these
questions.

So, it is a natural question to ask whether General Relativity can be
formulated with a polynomial Lagrangian. For pure vacuum General
Relativity, the answer is Yes and one such formulation is provided by
the cubic Einstein-Palatini formulation~\cite{Einstein,Palatini,Deser}
\begin{equation}\label{eq:EP}
	L_{\u a}[\g,C] = \g_{\u a}^{bc} (\bar{R}_{bc} + R_{bc}[C]) ,
	\quad\text{with}\quad 
	R_{bc}[C] = - \bnabla_b C^{d}_{dc} + \bnabla_d C^d_{bc}
		+ C^d_{bc} C_d - C^e_{bd} C^d_{ce} ,
\end{equation}
where we should interpret the inverse densitized metric $\g^{ab}_{\u a}
= v_{\u a}[g] g^{bc}$ as a fundamental variable, along with the
Christoffel tensor $C^d_{bc}$, which parametrizes the difference between
a connection $\nabla$ and the Levi-Civita connection $\bnabla$ for a
given background metric $\bar{g}$. Thus, $\bar{R}_{bc}$ is the Ricci
tensor of $\bar{g}$ and $\bar{R}_{bc} + R_{bc}[C]$ becomes equal to the
Ricci tensor for $g$, once $C^d_{bc}$ assumes the standard Levi-Civita
form for $g$. Since the variation with respect to $C$ yields an
equation for $C$ equivalent to $\nabla_d g_{bc} = 0$, eliminating the
auxiliary Christoffel tensor from the Lagrangian gives back the standard
Einstein-Hilbert form, though written in terms of $\g$ and thus commonly
known as the Goldberg Lagrangian~\cite{Goldberg}.

Unfortunately, it is not possible to maintain polynomial form with the
same variables as above when expanding the Lagrangian by including a
cosmological constant term or couplings to matter fields, both of which
are of interest in physical applications. The challenge then becomes to
introduce new auxiliary fields, like Lagrange multipliers, in terms of
which a sufficiently rich Lagrangian may be written in polynomial form.
We will content ourselves with adding (a) a cosmological constant, (b) a
massive scalar field, (c) a (Maxwell or Yang-Mills) gauge field, (d) a
Dirac spinor field, and (e) gauge-fixing and corresponding ghost terms
for those fields that have gauge symmetries. Point (e) is a necessary
ingredient for a consistent covariant quantization of this field
theory~\cite{HT}. Below, we give the final result, with some
explanations:

The field content is $\Phi = (\g,C,B,u,\bar{u}, v,w, \phi,
\hat{A},\hat{F},\hat{b},\hat{z},\hat{\bar{z}}, e,f,S,T, \psi)$, with its
interpretation below. The hatted fields are Lie algebra valued. The Lie
algebra is semi-simple, with commutator $[-,-]$ and invariant positive
definite inner product $\langle-,-\rangle$. The real scalar and complex
spinor multiplets carry representations of the Lie algebra, denoted
$(-)\cdot\phi$ or $(-)\cdot\psi$, and invariant inner products
$\langle-,-\rangle$, with $\langle\phi,A\cdot\phi\rangle = -\langle A
\cdot \phi, \phi\rangle$ and $\langle \psi, A\cdot\psi \rangle = \langle
A\cdot\psi, \psi \rangle$. We denote by $\bar{\gamma}^a$ a given choice
of hermitian $\gamma$-matrices for the given background metric
$\bar{g}_{bc}$, with $\bnabla$ its Levi-Civita connection, extended to
spinors such that $\bnabla_b \bar{\gamma}^c = 0$.

\begin{equation*}
\begin{aligned}
	\g_{\u a}^{bc} &\colon
		\text{inverse densitized metric,}
	&
	\hat{A}_b &\colon
		\text{gauge vector field,}
	\\
	C^a_{(bc)} &\colon
		\text{Christoffel tensor,}
	&
	\hat{F}_b^c &\colon
		\text{gauge curvature tensor,}
	\\
	B_c &\colon
		\text{de~Donder Nakanishi-Lautrup field,}
	&
	\hat{b} &\colon 
		\text{Lorenz Nakanishi-Lautrup field,}
	\\
	u^b, \bar{u}_c &\colon
		\text{diffeomorphism ghost and anti-ghost,}
	&
	\hat{z}, \hat{\bar{z}} &\colon
		\text{gauge ghost and anti-ghost,}
	\\
	\phi &\colon
		\text{scalar field (multiplet),}
	&
	\psi &\colon
		\text{spinor (multiplet),}
	\\
	v_{\u a} &\colon
		\text{metric volume form,}
	&
	w^{\u{k}^4\cdots\u{k}^n}&\colon
		\text{its Lagrange multiplier,}
	\\
	e_{(bc)} &\colon
		\text{Lorentz frame field,}
	&
	f^{(bc)} &\colon
		\text{its Lagrange multiplier,}
	\\
	T_{bcd} &\colon
		\text{spin torsion,}
	&
	S^{bcd}&\colon
		\text{its Lagrange multiplier.}
\end{aligned}
\end{equation*}

\begin{multline}
	L_{\u a}[\Phi]
	= \underbrace{\g_{\u a}^{bc} (\bar{R}_{bc} + R_{bc}[C])}_{(a1):3}
		-\underbrace{B_b (\g_{\u a}^{bc} B_c - 2 \bnabla_c \g_{\u a}^{bc})}_{(a2):3}
		-\underbrace{2\bnabla_{(b} \bar{u}_{c)} (\Lie_u \g)^{bc}_{\u a}}_{(a3):3}\\
	+ \underbrace{w^{\u{k}^4\cdots \u{k}^n}(G_{\u{a}\u{k}^4\cdots\u{k}^n}[\g]
			+ n!v_{\u{a}}v_{\u{k}^4}\cdots v_{\u{k}^n})}_{(b0):n+1} 
	- \underbrace{\frac{1}{2} (\g_{\u a}^{bc} \langle(\d\phi)_b,(\d\phi)_c\rangle
			+ m_\phi^2 \langle \phi,\phi\rangle v_{\u a})}_{(b1):3} \\
	+ \underbrace{\frac{1}{4} \langle \hat{F}_b^c, v_{\u a} \hat{F}_c^b - 2\g_{\u a}^{bd} (D\hat{A})_{cd}\rangle}_{(c1):3+1}
		+ \underbrace{\frac{1}{2} \langle \hat{b}, v_{\u a} \hat{b} - 2\g_{\u a}^{bc} \bnabla_b \hat{A}_c \rangle}_{(c2):3}
		+ \underbrace{\frac{1}{2} \langle \bnabla_b (\g_{\u a}^{bc} \hat{\bar{z}}), (D\hat{z})_{c} \rangle}_{(c3):3+1} \\
	+ \underbrace{f^{de} (\g_{\u a}^{bc} e_{bd} e_{ce} - v_{\u a} \bar{g}_{de})
			+ S^{bcd} [v_{\u a} e_{cc'} \bar{g}^{c'c''} T_{bc''d}
				- (e_{fb} \bnabla_c e_{de}) \g^{ef}_{\u a}]}_{(d0):4} \\
		\underbrace{- \frac{1}{2} \g^{bd}_{\u a} e_{bc}
			(\langle \psi , i \bar{\gamma}^c \bnabla_d \psi \rangle
				+ \langle i \bar{\gamma}^c \bnabla_d \psi, \psi \rangle)
			- \frac{1}{4} v_{\u a} \langle \psi,
					i \bar{\gamma}^{[b} \bar{\gamma}^{c} \bar{\gamma}^{d]}
					T_{bcd} \psi \rangle
		- m_\psi \langle \psi, \psi \rangle v_{\u a}}_{(d1):4} \\
	- \underbrace{2\Lambda v_{\u a}}_{(aa):1}
	- \underbrace{\lambda \phi^4 v_{\u a}}_{(bb):5}
	- \underbrace{\frac{m_A^2}{2} \g_{\u a}^{bc} \langle \hat{A}_b , \hat{A}_c \rangle}_{(cc):3}
	+ \underbrace{\frac{q_\phi}{2} \g_{\u a}^{bc}
		(\langle \bnabla_c\phi , \hat{A}_b\cdot\phi \rangle
		+ \langle \hat{A}_c\cdot\phi , \bnabla_b \phi \rangle)
		+ \frac{q_\phi^2}{2} \g_{\u a}^{bc} \langle \hat{A}_b\cdot \phi, \hat{A}_c\cdot \phi \rangle}_{(bc):5} \\
	- \underbrace{\mu \phi \langle \psi, \psi \rangle v_{\u a}}_{(bd):4}
	- \underbrace{\frac{q_\psi}{2} \g_{\u a}^{bd} e_{bc}
		(\langle \psi, \bar{\gamma}^c \hat{A}_d \cdot \psi \rangle
		+ \langle \bar{\gamma}^c \hat{A}_d \cdot \psi, \psi \rangle)}_{(cd):5}
	,
\end{multline}
where the terms are labeled by their role and polynomial degree, and
we have also used
\begin{align}
	R_{bc}[C]
		&= -\bnabla_b C^{d}_{dc} + \bnabla_d C^d_{bc}
			+ C^d_{bc} C_d - C^e_{bd} C^d_{ce} , \\
	(\Lie_u \g)^{bc}_{\u a}
		&= u^d \bnabla_d \g_{\u a}^{bc} - 2 \g_{\u a}^{d(b}\bnabla_d u^{c)}
			+ \g_{\u a}^{bc} \bnabla_d u^d , \\
	G_{\u{a}\u{k}^4\cdots\u{k}^n}[\g]
		&= \g^{b_1c_1}_{\u{a}} \g^{b_2c_2}_{\u{b}}
			\g^{b_3c_3}_{\u{c}} \g^{b_4c_4}_{\u{k}^4} \cdots \g^{b_nc_n}_{\u{k}^n}
			\quad \text{(recalling $\u{b} = b_1\cdots b_n$, $\u{c} = c_1\cdots c_n$),}
		\\
	(D\hat{A})_{dc}
		&= (\d \hat{A})_{dc} + [\hat{A}_d,\hat{A}_c] , \\
	(D\hat{z})_{c}
		&= (\d \hat{z})_c + [\hat{A}_c,\hat{z}] .
\end{align}
When $\hat{A}$ is valued in an abelian Lie algebra, the polynomial
degrees of $(c1),(c3)$ drop by $1$.

The roles of the various terms are as follows.
\begin{equation*}
\begin{aligned}
	(a1) &\colon
		\text{gravity kinetic term,}
	&
	(c1) &\colon
		\text{gauge kinetic term,}
	\\
	(a2) &\colon
		\text{de Donder gauge-fixing term,}
	&
	(c2) &\colon
		\text{Lorenz gauge-fixing term,}
	\\
	(a3) &\colon
		\text{diffeomorphism ghost kinetic term,}
	&
	(c3) &\colon
		\text{gauge ghost kinetic term,}
	\\
	(b0) &\colon
		\text{auxiliary fields for scalars,}
	&
	(d0) &\colon
		\text{auxiliary fields for spinors,}
	\\
	(b1) &\colon
		\text{scalar kinetic term with mass,}
	&
	(d1) &\colon
		\text{spinor kinetic term with mass,}
	\\
	(aa) &\colon
		\text{cosmological constant term,}
	&
	(bc) &\colon
		\text{scalar-gauge coupling,}
	\\
	(bb) &\colon
		\text{scalar potential,}
	&
	(bd) &\colon
		\text{(Yukawa) scalar-spinor coupling,}
	\\
	(cc) &\colon
		\text{(Proca) vector mass term,}
	&
	(cd) &\colon
		\text{gauge-spinor coupling.}
\end{aligned}
\end{equation*}
It should be noted that the (Proca) vector mass term with $m_A^2 \ne 0$
is incompatible with the vector gauge invariance. So, for consistency,
either only $(cc)$ or only the $(c2)$ and $(c3)$ terms should be
included. The scalar and spinor masses, $m_\phi^2$ and $m_\psi$, the
scalar coupling, $\lambda$, the scalar and spinor charges, $q_\phi$ and
$q_\psi$, and the Yukawa coupling, $\mu$, should be thought of as
tensors with respect to the appropriate multiplet structures.

Upon eliminating all the auxiliary fields (those that can be eliminated
by solving their Euler-Lagrange equations algebraically), the resulting
Lagrangian is the standard Lagrangian for General Relativity coupled to
a Standard Model-like matter theory. The overall polynomial degree of
the Lagrangian is $\max\{5, n+1\}$, with the dimension dependent degree
coming from the definition of the auxiliary volume form field, $v_{\u
a}$. The above form is far from the only way of putting a similar
Lagrangian in polynomial form (cf.~\cite{Tseytlin,Katanaev}). It is
interesting to consider whether a polynomial form could be achieved with
a smaller number of auxiliary fields or smaller polynomial degree in
various terms.

\bibliographystyle{utphys-alpha}
\bibliography{paper-lepage}

\end{document}